\documentstyle[12pt]{article}

\newcommand {\e} {\mbox{\rm e}}

\newcommand {\nn}    {\nonumber}
\newcommand {\vs}[1]  { \vspace*{#1 cm} }

\newcounter{eq}
\newcounter{sc}

\newcommand {\MPL}  {Mod.Phys.Lett.}
\newcommand {\NP}   {Nucl.Phys.}
\newcommand {\PL}   {Phys.Lett.}
\newcommand {\PR}   {Phys.Rev.}
\newcommand {\PRL}   {Phys.Rev.Lett.}

\newcommand {\AP}   {Ann.of Phys.}

\def\overleftrightarrow#1{\vbox{\ialign{##\crcr
 $\leftrightarrow$\crcr\noalign{\kern-1pt\nointerlineskip}
 $\hfil\displaystyle{#1}\hfil$\crcr}}}

\newlength{\minitwocolumn}
\begin{document}

\begin{flushright}
EDO-EP-39\\
March, 2001\\
\end{flushright}
\vspace{30pt}

\pagestyle{empty}
\baselineskip15pt

\begin{center}
{\large\bf Kaluza-Klein Reduction in a Warp Geometry Revisited

 \vskip 1mm
}

\vspace{20mm}

Ichiro Oda
          \footnote{
          E-mail address:\ ioda@edogawa-u.ac.jp
                  }
\\
\vspace{10mm}
          Edogawa University,
          474 Komaki, Nagareyama City, Chiba 270-0198, JAPAN \\

\end{center}


\vspace{15mm}
\begin{abstract}
We study the Kaluza-Klein dimensional reduction of zero-modes of
bulk antisymmetric tensor fields on a non-compact extra dimension 
in the Randall-Sundrum model.
It is shown that in the Kaluza-Klein reduction on a non-compact 
extra dimension we have in general a zero-mode depending on a 
fifth dimension in addition to a conventional constant zero-mode
in the Kaluza-Klein reduction on a circle. 
We examine the localization property of these zero-modes on a flat
Minkowski 3-brane. In particular, it is shown that a 2-form and a 3-form 
on the brane can be respectively obtained from a 3-form and 4-form in 
the bulk by taking the zero-mode dependent on the fifth dimension.   

\vspace{15mm}

\end{abstract}

\newpage
\pagestyle{plain}
\pagenumbering{arabic}

\rm

The gravity-localized models in a brane world have the intriguing
feature that even if there is a non-compact extra dimension,
the graviton is sharply localized on a flat Minkowski brane,
thereby reproducing the four-dimensional Newton's law with
negligible corrections on the brane \cite{Randall1, Randall2}.
(For multi-brane models, see \cite{Oda1}.) 
It is then natural to ask whether matter and gauge fields
in addition to the graviton are also localized on the brane by a
gravitational interaction. If the entire local fields are 
trapped on the brane, we could regard such a 3-brane as a candidate of 
our real world. Indeed, in superstring theory, matter and gauge fields 
are naturally confined to D3 branes due to open strings ending on the 
branes while the gravity is free to propagate in a bulk space-time due to 
closed strings living in the bulk \cite{Pol}.
On the other hand, in local field theory, it has been well known that 
except a 1-form gauge field  all the local fields are also localized 
on a brane by a gravitational interaction \cite{Dvali, Pomarol, Rizzo, 
Bajc, Grossman, Oda2}. (For a review see \cite{Oda3}.)

Recently, we have proposed a completely new mechanism for trapping 
a 1-form bulk gauge field on a flat brane in the gravity-localized model
\cite{Oda4}. 
The key idea is to consider a topological term and a 3-form  
action in addition to the $U(1)$ gauge field action in order to give 
the mass term with a 'kink' profile to the bulk gauge field. 
(This mass generation mechanism is called 'topological mass generation'
or 'topological Higgs mechanism' \cite{Schonfeld, Deser, Oda5}.)
This new localization mechanism is very similar to that of fermions
\cite{Jackiw} in the sense that in both the cases the zero-modes share 
the same exponentially damping form and then some inequality between the 
bulk mass and the constant in the warp factor is needed to insure the 
localization and massless condition of brane gauge field at the same time.

Accordingly, it has been now shown that the entire bulk fields ranging 
from a spin-0 scalar to a spin-2 graviton are localized on a flat brane only
by a gravitational interaction in the Randall-Sundrum model \cite{Randall1, 
Randall2}. 
The fields we have left aside in the above study are the antisymmetric 
tensor fields. Since many antisymmetric fields appear in the low energy
effective action of superstring theory, we should also study the localization 
property of these fields on a flat 3-brane.
Recently, it has been shown that a 2-form potential is also
confined to a flat brane in the Randall-Sundrum model \cite{Duff}. 
(See also closely related papers \cite{Chris}.)
The important observation there is that a 3-form in a bulk yields
a 2-form on a brane by the nontrivial Kaluza-Klein dimensional reduction,
which has been found through the Hodge duality relation between a massless 
0-form and a massless 3-form in the five-dimensional space-time. Recall 
that such a nontrivial zero mode depending on a fifth dimension has also been 
utilized in the study of the localization of bulk fields in the
locally-localized 
gravity models \cite{Oda6}.

The modest aim of the present paper is to clarify the Kaluza-Klein 
reduction of antisymmetric tensor fields on a non-compact extra dimension
in the Randall-Sundrum model.
We shall make use of a more general approach rather than an approach on the 
basis of the Hodge duality \cite{Duff} since the latter approach relies 
heavily relies on the Hodge duality, which holds only on-shell. 
Because of it, for instance, provided that there are 
mass terms and topological terms such as Chern-Simons and BF terms,
only the former approach provides us with a useful method. Actually, such
a situation has been already appeared in Ref. \cite{Oda4}.
For completeness, in this paper, we shall consider the entire antisymmetric
tensor fields ranging from a 0-form scalar field to a 4-form potential 
in the five-dimensional Randall-Sundrum model. Incidentally the
generalization 
to the non-abelian gauge fields and higher dimensions would be
straightforward.

We shall start by fixing our model setup.  
The metric ansatz we take is of the Randall-Sundrum form \cite{Randall2}:
\begin{eqnarray}
ds^2 &=& g_{MN} dx^M dx^N  \nn\\
&=& \e^{-A(r)} \eta_{\mu\nu} dx^\mu dx^\nu + dr^2,
\label{1}
\end{eqnarray}
where $M, N, \cdots$ are five-dimensional space-time indices and 
$\mu, \nu \cdots$ are four-dimensional brane indices. The metric on the 
brane $\eta_{\mu\nu}$ denotes the four-dimensional flat Minkowski metric 
with signature $(-,+,+,+)$. Moreover, $A(r) = 2 k |r|$ where $k$ is a 
positive constant and the fifth dimension $r$ runs from $-\infty$ to 
$\infty$.
We have a model setup in mind where a single flat 3-brane sits
at the origin $r = 0$ of the fifth dimension and various antisymmetric 
tensor fields reside in a bulk. 
We will assume that the background metric is not 
modified by the presence of the bulk fields, that is, we will neglect 
the back-reaction on the metric from the bulk fields.
Under such a model setup, we look for zero-modes of the bulk fields 
in a simple Kaluza-Klein ansatz such that the zero-modes are 
not only normalizable but also localized sharply near the brane. 
Here it is worthwhile to stress one important point 
that the normalizable condition, which is equivalent to the convergence 
of the integral over the fifth dimension $r$ in front of the kinetic terms, 
is usually thought to be a necessary and sufficient condition 
for the localization of the bulk fields on a brane \cite{Bajc}.
However, as shown in Refs. \cite{Oda4, Oda6}, we sometimes meet the 
situation where the zero-modes are normalizable but spread rather widely 
in a bulk. Perhaps such the widely spread zero-modes would be in
contradiction 
with experimental results such as the charge conservation law. 
Thus, in order to show a complete localization of
bulk fields on a brane, we have to check that the normalized zero-modes in a 
flat space take an exponentially damping form in addition to the normalizable 
condition.

We shall start with a massless 0-form real scalar field \cite{Bajc}. 
The action of a 0-form potential $\Phi$ is given by
\begin{eqnarray}
S_0 = - \frac{1}{2} \int d^5 x \sqrt{-g}  g^{M N} \partial_M \Phi
\partial_N \Phi.
\label{2}
\end{eqnarray}
Then the equation of motion becomes
\begin{eqnarray}
\frac{1}{\sqrt{-g}} \partial_M \left(\sqrt{-g} g^{M N} \partial_N 
\Phi \right) = 0.
\label{3}
\end{eqnarray}
As mentioned before, we shall look for a zero-mode solution with the 
following simple Kaluza-Klein ansatz  
\begin{eqnarray}
\Phi(x^M) = \phi(x^\mu) u(r),
\label{4}
\end{eqnarray}
where we assume the equation of motion for the brane scalar field, 
$\Box \phi = 0$ with being $\Box \equiv \eta^{\mu\nu}
\partial_\mu \partial_\nu$.

With the ansatz (\ref{4}), Eq. (\ref{3}) reduces to a single differential
equation for $u(r)$:
\begin{eqnarray}
\partial_r \left( \e^{- 2 A(r)} \partial_r u \right) = 0.
\label{5}
\end{eqnarray}
The general solution to this equation is easily found to be
\begin{eqnarray}
u(r) = \frac{u_1}{4 k} \e^{2 A(r)}  + u_0,
\label{6}
\end{eqnarray}
where $u_0$ and $u_1$ are integration constants. To derive this solution,
we have assumed that $\partial_r u(0) = 0$, which would stem from
the fact that if we impose $Z_2$ orbifold symmetry on the fifth
dimension, the self-adjointness of the differential operator requires
the Neumann boundary condition at $r = 0$. Henceforth we will also assume
$\partial_r u(0) = 0$. 

Here we wish to emphasize an important fact associated with the 
Kaluza-Klein reduction on a non-compact dimension in a warp geometry, 
which will indeed play a critical role in deriving
zero-modes bound to a brane from fields in a bulk. 
In the ordinary Kaluza-Klein dimensional reduction scenario on a 
compact circle, a zero-mode is just a constant since there is a cyclic 
symmetry $r \rightarrow r + 2 \pi$ and usually a zero-mode respecting 
this symmetry is only a constant. (Of course, the excited modes have
a factor $\e^{i n r}$ with $n$ being integers, which obviously
respects the cyclic symmetry.) On the other hand, in a non-compact case
at hand, we have no room to impose such a symmetry once the position of
a brane is fixed, so that a zero-mode with the nontrivial dependence
on an extra dimension is allowed.

Plugging the ansatz (\ref{4}) into the starting action (\ref{2}),
the action can be cast to the form
\begin{eqnarray}
S_0^{(0)} = \int d^4 x dr \Big[ - \frac{1}{2} \partial^\mu \phi 
\partial_\mu \phi \e^{- A(r)} u^2(r) - \frac{1}{2} \phi^2 \e^{- 2 A(r)} 
(\partial_r u)^2 \Big].
\label{7}
\end{eqnarray}
Here we take a special solution $u(r) = u_0$, which is a constant
zero-mode, since the solution has a more convergent character than
the solution with the nonzero $u_1$. Then, the integral over $r$
in front of the kinetic term in Eq. (\ref{7}) is found to be finite
as follows:
\begin{eqnarray}
I \equiv \int_{- \infty}^{\infty} dr \e^{- A(r)} u^2(r)
= \frac{u_0^2}{k}. 
\label{8}
\end{eqnarray}
This convergence of the $r$-integral implies the normalizability
of the zero mode $u(r) = u_0$ of a bulk scalar field.
In order to show a sharp localization, it is necessary to check that
the normalized zero-mode has an exponentially damping form \cite{Oda4}.
Actually, the normalized zero-mode in a flat space has the form
\begin{eqnarray}
\hat{u}(r) = \frac{1}{\sqrt{I}} \e^{-\frac{1}{2} A(r)} u(r)
= \sqrt{k} \e^{- k |r|},
\label{9}
\end{eqnarray}
which obviously indicates that the brane scalar field is sharply
localized near the brane sitting at $r = 0$ as long as $k \gg 1$. 

Next, we shall turn to a massless 1-form potential, that is, a 
$U(1)$ gauge field. The path of arguments is very similar to the case 
of a 0-form potential. The action is 
\begin{eqnarray}
S_1 = -\frac{1}{4} \int d^5 x \sqrt{-g}  g^{M N} g^{R S} F_{M R} F_{N S}, 
\label{12}
\end{eqnarray}
where $F_{MN} = 2 \partial_{[M} A_{N]} = \partial_M A_N - \partial_N A_M$. 
The equations of motion read 
\begin{eqnarray}
\frac{1}{\sqrt{-g}} \partial_M \left(\sqrt{-g} g^{M N} g^{R S} F_{N S}
\right) = 0. 
\label{13}
\end{eqnarray}
We search for a zero-mode solution of the form
\begin{eqnarray}
A_\mu(x^M) &=& a_\mu(x^\lambda) u(r), \nn\\
A_r(x^M) &=& a(x^\lambda) v(r),
\label{14}
\end{eqnarray}
where we assume the equations of motion on a brane,
$\partial^\nu f_{\mu\nu} = \Box a = 0$ with the definition of 
$f_{\mu\nu} = \partial_\mu a_\nu -  \partial_\nu a_\mu$.
(Here we have not fixed the gauge symmetries in five dimensions.)

With this ansatz, Eq. (\ref{13}) reduces to two differential
equations for $u(r)$ and $v(r)$:
\begin{eqnarray}
\partial_r \left( \e^{- A(r)} \partial_r u \right) &=& 0, \nn\\
\partial_r \left( \e^{- A(r)} v \right) &=& 0,
\label{15}
\end{eqnarray}
whose general solution is given by
\begin{eqnarray}
u(r) &=& \frac{u_1}{2 k} \e^{A(r)} + u_0, \nn\\
v(r) &=& v_0 \e^{A(r)}.
\label{16}
\end{eqnarray}
where $u_0$, $u_1$  and $v_0$ are integration constants. From 
the better convergent property, we shall choose the solution
with $u_1 = 0$ in what follows.

Plugging this solution into the classical action (\ref{12}), 
the action takes the form
\begin{eqnarray}
S_1^{(0)} = \int d^4 x dr \Big[ - \frac{1}{4} f^2_{\mu\nu} u^2_0
- \frac{1}{2} \partial_\mu a \partial^\mu a \e^{A} v^2_0 \Big]. 
\label{17}
\end{eqnarray}
Note that the two $r$-integrals in front of the vector and scalar
kinetic terms, those are, 
$I_1 \equiv \int_{- \infty}^{\infty} dr u^2_0$ and
$I_2 \equiv \int_{- \infty}^{\infty} dr \e^A v^2_0$ diverge,
thereby implying that a 1-form $a_\mu$ and a 0-form $a$
are not localized on a brane, which is the well-known fact.
This problem has been recently circumvented by coupling the bulk gauge
field to a 3-form potential through a topological term \cite{Oda4}. 

We are now ready to consider a 2-form potential, in other words, the
Kalb-Ramond second-rank antisymmetric tensor field. In five dimensions, 
a 2-form is dual to a 1-form, so it is expected 
that a 2-form is also trapped on a brane as in a 1-form.
However, it is impossible to derive the 2-form confined to
the brane from the conventional action of a 2-form. We are now familiar 
with the fact that such a trapped 2-form can be obtained from the action of
a 3-form through the Kaluza-Klein reduction on a non-compact extra
dimension \cite{Duff}. 

The classical action of a massless 2-form is given by 
\begin{eqnarray}
S_2 &=& - \frac{1}{12} \int d^5 x \sqrt{-g}  g^{M_1 N_1} g^{M_2 N_2} 
g^{M_3 N_3} F_{M_1 M_2 M_3} F_{N_1 N_2 N_3}, 
\label{18}
\end{eqnarray}
where $F_{MNP} = 3 \partial_{[M} A_{NP]} = \partial_M A_{NP} + 
\partial_N A_{PM} + \partial_P A_{MN}$. The equations of motion
are 
\begin{eqnarray}
\frac{1}{\sqrt{-g}} \partial_{M_3} \left(\sqrt{-g} g^{M_1 N_1} g^{M_2 N_2} 
g^{M_3 N_3} F_{N_1 N_2 N_3} \right) = 0. 
\label{19}
\end{eqnarray}
The natural Kaluza-Klein ansatz for a zero-mode solution is 
\begin{eqnarray}
A_{\mu\nu}(x^M) &=& a_{\mu\nu}(x^\lambda) u(r), \nn\\
A_{r \mu}(x^M) &=& a_\mu(x^\lambda) v(r),
\label{20}
\end{eqnarray}
where the following equations on a brane are imposed: 
$\partial^\rho f_{\mu\nu\rho} = \partial^\nu f_{\mu\nu} = 0$ 
with the definition being $f_{\mu\nu\rho} = 3 \partial_{[\mu} a_{\nu\rho]}$.

With this ansatz, Eq. (\ref{19}) reduces to differential equations:
\begin{eqnarray}
\partial_r^2 u &=& 0, \nn\\
\partial_r v &=& 0,
\label{21}
\end{eqnarray}
whose general solution is simply given by 
\begin{eqnarray}
u(r) &=& u_1 r + u_0, \nn\\
v(r) &=& v_0.
\label{22}
\end{eqnarray}
Let us select a solution with $u_1 = 0$ because of the better convergent 
property.

Substituting this solution into the action (\ref{18}), the action 
reduces to the form
\begin{eqnarray}
S_2^{(0)} = \int d^4 x  dr \Big[ - \frac{1}{12} f^2_{\mu\nu\rho} \e^A u^2_0
- \frac{1}{4} f^2_{\mu\nu} v^2_0 \Big]. 
\label{23}
\end{eqnarray}
Note that the two integrals over $r$, those are, 
$I_1 \equiv \int_{- \infty}^{\infty} dr \e^A u^2_0$ and
$I_2 \equiv \int_{- \infty}^{\infty} v^2_0$ diverge,
so that neither a 2-form $a_{\mu\nu}$ nor a 1-form $a_\mu$ are localized 
on a brane. Therefore, starting with the action of a 2-form in a bulk
we cannot obtain a 2-form bound to a brane.

Here let us comment on the possibility of generalizing a new mechanism 
for trapping of a 1-form to the case of a 2-form \cite{Oda4}.
Perhaps, the most plausible possibility would be to consider the
following action:
\begin{eqnarray}
S_2 = \int \Big[ - \frac{1}{2} B_2 \wedge * B_2 - \frac{1}{2} C_2 
\wedge * C_2 + m B_2 \wedge d C_2 \Big], 
\label{A}
\end{eqnarray}
where $B_2$ and $C_2$ are two 2-form fields, and we have used the
form notations for convenience. With the gauge conditions
$B_{r M} = C_{r M} =0$, we have analyzed the localization property
of $B_{\mu\nu}$ and $C_{\mu\nu}$. Unfortunately, it turns out that
these two 2-forms have normalizable zero-modes but are not localized
sharply on a flat 3-brane, as in the locally-localized gravity
models \cite{Oda6}.  

Let us turn our attention to a 3-form potential.
We will see that a 3-form in a bulk yields a 2-form trapped
on a brane via the Kaluza-Klein reduction.
The classical action of a massless 3-form potential reads  
\begin{eqnarray}
S_3 &=&  -\frac{1}{48} \int d^5 x \sqrt{-g}  g^{M_1 N_1} g^{M_2 N_2} 
g^{M_3 N_3} g^{M_4 N_4} F_{M_1 M_2 M_3 M_4} F_{N_1 N_2 N_3 N_4}, 
\label{24}
\end{eqnarray}
where $F_{MNPQ} = 4 \partial_{[M} A_{NPQ]} = \partial_M A_{NPQ} - 
\partial_N A_{MPQ} + \partial_P A_{MNQ} - \partial_Q A_{MNP}$. 
The equations of motion are then 
\begin{eqnarray}
\frac{1}{\sqrt{-g}} \partial_{M_4} \left(\sqrt{-g} g^{M_1 N_1} g^{M_2 N_2} 
g^{M_3 N_3} g^{M_4 N_4} F_{N_1 N_2 N_3 N_4} \right) = 0. 
\label{25}
\end{eqnarray}
We shall make the Kaluza-Klein ansatz  
\begin{eqnarray}
A_{\mu\nu\rho}(x^M) &=& a_{\mu\nu\rho}(x^\lambda) u(r), \nn\\
A_{r \mu\nu}(x^M) &=& a_{\mu\nu}(x^\lambda) v(r), 
\label{26}
\end{eqnarray}
where we assume that $\partial^\sigma f_{\mu\nu\rho\sigma} 
= \partial^\rho f_{\mu\nu\rho} = 0$ with the definition being 
$f_{\mu\nu\rho\sigma} = 4 \partial_{[\mu} a_{\nu\rho\sigma]}$.

Then Eq. (\ref{25}) becomes
\begin{eqnarray}
\partial_r \left( \e^A \partial_r u \right) &=& 0, \nn\\
\partial_r \left( \e^A v \right) &=& 0.
\label{27}
\end{eqnarray}
The general solution is given by 
\begin{eqnarray}
u(r) &=& - \frac{u_1}{2 k} \e^{-A} + u_0, \nn\\
v(r) &=& v_0 \e^{-A}.
\label{28}
\end{eqnarray}

Inserting the ansatz (\ref{26}) to the starting action (\ref{24}),
we have
\begin{eqnarray}
S_3^{(0)} = \int d^4 x dr \Big[ - \frac{1}{48} f^2_{\mu\nu\rho\sigma} 
\e^{2A} u^2 - \frac{1}{12} ( a_{\mu\nu\rho} \partial_r u - 
f_{\mu\nu\rho} v )^2 \e^A  \Big]. 
\label{29}
\end{eqnarray}
We shall discuss the two cases separately, one of which is $u_1 = 0$
and the other is $u_1 \neq 0$.

In the case of $u(r) = u_0$ (i. e., $u_1 = 0$), the above action takes
the form
\begin{eqnarray}
S_3^{(0)} = \int d^4 x dr \Big[ - \frac{1}{48} f^2_{\mu\nu\rho\sigma} 
\e^{2A} u^2_0 - \frac{1}{12} f^2_{\mu\nu\rho} \e^A v^2 \Big]. 
\label{30}
\end{eqnarray}
The integral over $r$ in front of the kinetic term for 
a 3-form $a_{\mu\nu\rho}$, $I_1 \equiv \int_{- \infty}^{\infty} dr 
\e^{2A} u^2_0$ diverges, so the 3-form is not localized on a brane.
On the other hand, the integral in front of the kinetic term 
for a 2-form $a_{\mu\nu}$, $I_2 \equiv \int_{- \infty}^{\infty} \e^A v^2$ 
takes the finite value of $\frac{v^2_0}{k}$. This fact suggests that
the 2-form might be localized near the brane. Indeed, the normalized 
zero-mode for the 2-form in a flat space is given by
\begin{eqnarray}
\hat{v}(r) = \frac{1}{\sqrt{I_2}} \e^{\frac{1}{2} A(r)} v(r)
= \sqrt{k} \e^{- k |r|},
\label{31}
\end{eqnarray}
which shows that the 2-form $a_{\mu\nu}$, which has been obtained from
a bulk 3-form via the Kaluza-Klein reduction, is sharply localized on a
brane as long as $k \gg 1$. 
Incidentally, the zero-mode for the 3-form has a behavior like
$\hat{u}(r) \sim \e^A u_0 = \e^{2 k |r|} u_0$, so the 3-form resides
in a bulk away from a brane.
Hence, effectively on a brane, we have an action
\begin{eqnarray}
S_3^{(0)} = - \frac{1}{12} \int d^4 x  f^2_{\mu\nu\rho}, 
\label{32}
\end{eqnarray}
where we have redefined as $\frac{v_0}{\sqrt{k}} a_{\mu\nu} \rightarrow
a_{\mu\nu}$. 

Next, let us consider the case $u(r) = - \frac{u_1}{2 k} \e^{-A} + u_0$
 (i. e., $u_1 \neq 0$). We shall note first that in the case
of $u(r) = - \frac{u_1}{2 k} \e^{-A} + u_0$, we can rewrite a term in
(\ref{29}) as follows:
\begin{eqnarray}
a_{\mu\nu\rho} \partial_r u - f_{\mu\nu\rho} v 
= ( a_{\mu\nu\rho} - \frac{v_0}{u_1} f_{\mu\nu\rho}) u_1 \e^{-A}.
\label{33}
\end{eqnarray}
With the field redefinition (or equivalently, the gauge transformation)
\begin{eqnarray}
a_{\mu\nu\rho} \rightarrow a_{\mu\nu\rho} + \frac{v_0}{u_1} f_{\mu\nu\rho},
\label{35}
\end{eqnarray}
the action transforms as
\begin{eqnarray}
S_3^{(0)} = \int d^4 x dr \Big[ - \frac{1}{48} f^2_{\mu\nu\rho\sigma} 
\e^{2A} u^2 - \frac{1}{12} a^2_{\mu\nu\rho} \e^A (\partial_r u)^2 \Big].
\label{36}
\end{eqnarray}
In this action, the integral over $r$ in front of the kinetic term
obviously diverges, so that a 3-form $a_{\mu\nu\rho}$ is not localized
on a brane, but resides in a bulk. The real problem here is that this
time we have no a 2-form $a_{\mu\nu}$ on the brane, which should be
in contrast to the previous $u(r) = u_0$ case, where a 2-form stemming
from a 3-form is bound to the brane. To avoid this difficulty, Duff and
Liu have chosen $a_{\mu\nu\rho} = 0$ from the beginning although
their reasoning is completely different from ours and is based on
the Hodge duality \cite{Duff}. It is then natural to ask why the results
between ours and Duff et al. appear to be so different. The answer lies
in the gauge symmetries in the starting action (\ref{24}).
The action (\ref{24}) has the gauge symmetries
as well as off-shell reducible symmetries \cite{Batalin}: 
\begin{eqnarray}
\delta A_{MNP} &=& \partial_{[M} \varepsilon_{NP]}, \nn\\
\delta \varepsilon_{MN} &=& \partial_{[M} \varepsilon_{N]}, \nn\\
\delta \varepsilon_M &=& \partial_M \varepsilon,
\label{37}
\end{eqnarray}
whose number of degrees of freedom is $\frac{5 \times 2}{2} - 5 + 1 =6$, 
which exactly coincides with the number of dynamical degrees of freedom
involved in $A_{r\mu\nu}$. Accordingly, $a_{\mu\nu}$
can be gauge-fixed to be zero, which yields our result (\ref{36}), in other
words, no 2-form on a brane. From this viewpoint, the result by Duff et al.
is interpreted as follows: They have picked up the gauge conditions
$A_{\mu\nu\rho} = 0$ from the outset, so  $a_{\mu\nu}$ cannot be gauged away
any more and consequently appears in the final action (\ref{32}).

Finally, we would like to consider a massless 4-form field. This 
field is non-dynamical but it is of theoretical interest to check
if the approach at hand also applies to this case. It is expected
that such a non-dynamical field might play an important role in
the cosmological constant problem, so there would be also of some physical
importance.
The classical action of a massless 4-form  reads  
\begin{eqnarray}
S_4 = -\frac{1}{240} \int d^5 x \sqrt{-g} 
g^{M_1 N_1} g^{M_2 N_2} g^{M_3 N_3} g^{M_4 N_4} g^{M_5 N_5}
F_{M_1 M_2 M_3 M_4 M_5} F_{N_1 N_2 N_3 N_4 N_5}, 
\label{41}
\end{eqnarray}
where $F_{MNPQR} = 5 \partial_{[M} A_{NPQR]}$. 
The equations of motion then take the form
\begin{eqnarray}
\partial_{M_5} \left(\sqrt{-g} g^{M_1 N_1} g^{M_2 N_2} g^{M_3 N_3} 
g^{M_4 N_4} g^{M_5 N_5} F_{N_1 N_2 N_3 N_4 N_5} \right) = 0. 
\label{42}
\end{eqnarray}
Making the Kaluza-Klein ansatz  
\begin{eqnarray}
A_{\mu\nu\rho\sigma}(x^M) &=& a_{\mu\nu\rho\sigma}(x^\lambda) u(r), \nn\\
A_{r \mu\nu\rho}(x^M) &=& a_{\mu\nu\rho}(x^\lambda) v(r), 
\label{43}
\end{eqnarray}
with the assumption that $\partial^\sigma f_{\mu\nu\rho\sigma} 
= \partial^\sigma a_{\mu\nu\rho\sigma} = 0$, the action reads
\begin{eqnarray}
S_4^{(0)} = -\frac{1}{48} \int d^4 x dr  
(a_{\mu\nu\rho\sigma} \partial_r u - f_{\mu\nu\rho\sigma} v)^2
\e^{2A}. 
\label{44}
\end{eqnarray}
Moreover, Eq. (\ref{42}) becomes
\begin{eqnarray}
\partial_r \left( \e^{2A} \partial_r u \right) &=& 0, \nn\\  
\partial_r \left( \e^{2A} v \right) &=& 0.
\label{45}
\end{eqnarray}
The general solution is given by 
\begin{eqnarray}
u(r) &=& - \frac{u_1}{4 k} \e^{-2A} + u_0, \nn\\
v(r) &=& v_0 \e^{-2A}.
\label{46}
\end{eqnarray}

Again, we shall first consider the case of $u(r) = u_0$, that is,
$u_1 = 0$. Then, the action (\ref{44}) reduces to
\begin{eqnarray}
S_4^{(0)} = -\frac{1}{48} \int d^4 x dr  
f^2_{\mu\nu\rho\sigma} \e^{2A} v^2. 
\label{47}
\end{eqnarray}
Since $I \equiv \int_{- \infty}^{\infty} dr \e^{2A} v^2 = \frac{v^2_0}
{2 k}$, this zero-mode is a normalizable one. In fact, the normalized
zero-mode is given by $\hat{v}(r) = \sqrt{2 k} \e^{- 2 k |r|}$, so
a 3-form $a_{\mu\nu\rho}$ is localized on a brane.

Next, let us consider the case of $u(r) = - \frac{u_1}{4 k} \e^{-2A} 
+ u_0$. This time, the field redefinition
\begin{eqnarray}
a_{\mu\nu\rho\sigma} \rightarrow a_{\mu\nu\rho\sigma} 
+ \frac{v_0}{u_1} f_{\mu\nu\rho\sigma},
\label{48}
\end{eqnarray}
leads to 
\begin{eqnarray}
S_4^{(0)} &=& -\frac{1}{48} \int d^4 x dr  
a^2_{\mu\nu\rho\sigma} \e^{2A} (\partial_r u)^2 \nn\\
&=& -\frac{1}{48} \int d^4 x \frac{u_1^2}{2 k} a^2_{\mu\nu\rho\sigma}. 
\label{49}
\end{eqnarray}
Hence, as in a 3-form potential, in order to reproduce a 3-form on a
brane, we need to fix the gauge symmetries by $A_{\mu\nu\rho\sigma}
=0$. 

In conclusion, we have examined the Kaluza-Klein reduction on a non-compact
dimension of bulk antisymmetric tensor fields from a 0-form to a 4-form in
the Randall-Sundrum model. Compared with the ordinary Kaluza-Klein
reduction on a compact circle, we can find zero-modes which are manifestly
dependent on the extra dimension. We have seen that such nontrivial 
zero-modes provide us with a 2-form and a 3-form bound to a brane from
a 3-form and a 4-form in a bulk, respectively, through the Kaluza-Klein 
reduction. This observation has also been used in showing the localization
of various bulk fields in the locally-localized gravity models \cite{Oda6}.

\vs 1


\end{document}